\documentclass[preprint,5p]{elsarticle}
\usepackage{graphicx}
\usepackage{dcolumn}
\usepackage{bm}  
\usepackage{textcomp}  
\usepackage{amssymb}  
\usepackage{hyperref}

\hypersetup{colorlinks=true, urlcolor=blue, citecolor=blue}
\usepackage[displaymath]{lineno}
\usepackage{dblfloatfix}    % To enable figures at the bottom of page
\begin{document}

\title{\vspace{-15mm}\fontsize{24pt}{10pt}\selectfont\textbf{A Radial Time 
Projection Chamber for $\alpha$ detection in CLAS at JLab}}
\author[anl,ipn]{R.~Dupr\'e\corref{cor}}\ead{dupre@ipno.in2p3.fr} 
\author[jlab]{S.~Stepanyan} 
\author[anl,ipn]{M.~Hattawy} 
\author[anl,jlab]{N.~Baltzell} 
\author[anl]{K.~Hafidi} 
\author[infn]{M.~Battaglieri} 
\author[odu]{S.~Bueltmann} 
\author[infn]{A.~Celentano} 
\author[infn]{R.~De~Vita} 
\author[anl,utfsm]{A.~El~Alaoui} 
\author[msu]{L.~El~Fassi} 
\author[jlab]{H.~Fenker} 
\author[anl]{K.~Kosheleva} 
\author[odu]{S.~Kuhn} 
\author[infn]{P.~Musico} 
\author[infn]{S.~Minutoli} 
\author[uc]{M.~Oliver} 
\author[lpsc]{Y.~Perrin} 
\author[odu]{B.~Torayev}
\author[ipn,lpsc]{E.~Voutier} 

\address[anl]{Argonne National Laboratory, Argonne IL 60439, USA}
\address[ipn]{Institut de Physique Nucl\'eaire, CNRS/IN2P3 and Universit\'e Paris Sud, Orsay, France}
\address[jlab]{Jefferson Laboratory, Newport News, VA 230606, USA}
\address[infn]{INFN, Sezione di Genova, 16146 Genova, Italy}
\address[odu]{Old Dominion University, Norfolk, VA 23529, United States}
\address[utfsm]{Universidad T\'ecnica Federico Santa Mar\'ia, Casilla 110-V Valpara\'iso, Chile}
\address[msu]{Mississippi State University, Mississippi State, MS 39762-5167}
\address[uc]{University of Chicago, Chicago, IL 60637, United States}
\address[lpsc]{LPSC, Universit\'e Grenoble-Alpes, CNRS/IN2P3, Grenoble, France}

\cortext[cor]{Corresponding author}

\date{\today}

\begin{abstract}
A new Radial Time Projection Chamber (RTPC) was developed at the Jefferson 
Laboratory to track low-energy nuclear recoils to measure
exclusive nuclear reactions, such as coherent deeply virtual Compton scattering
and coherent meson production off $^4$He. In
2009, we carried out these measurements using the CEBAF Large
Acceptance Spectrometer (CLAS) supplemented by
the RTPC positioned directly around a gaseous $^4$He target, allowing a detection
threshold as low as 12~MeV for $^4$He. This article discusses the design,
principle of operation, calibration methods and performances of this RTPC.
\end{abstract}

\maketitle


\section{Introduction} \label{sec:level1}

\begin{figure}[tbp]
\centering \includegraphics[scale=0.3]{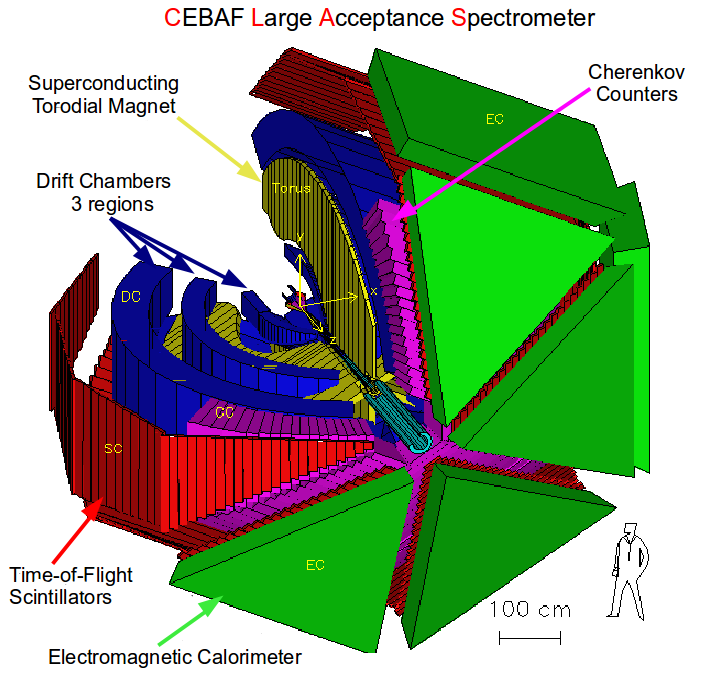}
\caption{A three dimensional representation of the baseline CLAS setup. The
   full description is given in the text.} \label{fig:CLAS}
\end{figure}

Until recently, the Thomas Jefferson National Accelerator Facility, in 
Newport News, Virginia, USA, has provided high power electron beams of 
up to 6 GeV energy and 100$\%$ duty factor to three experimental Halls (A, B, C) 
simultaneously. The CEBAF Large
Acceptance Spectrometer (CLAS)~\cite{Mecking:2003zu}, located
in Hall-B, was based on a superconducting toroidal magnet and composed of 
several sub-detectors. Figure~\ref{fig:CLAS} shows a three dimensional 
representation of the baseline CLAS spectrometer:
\begin{itemize}
 \item Three regions of Drift Chambers (DC) for the tracking of charged 
       particles~\cite{Mestayer:2000we}.
 \item Superconducting toroidal magnet to bend the trajectories 
       of charged particles, thus allowing momentum measurement with the DC tracking information.
 \item Threshold Cherenkov Counters (CC) for electron identification at momenta 
    $<2.7$ GeV/c~\cite{Adams:2001kk}.
 \item Scintillation Counters (SC) to identify charged hadrons by measuring their 
       time of flight~\cite{Smith:1999ii}.
 \item Electromagnetic Calorimeters (EC) for identification of electrons, 
       photons and neutrons~\cite{Amarian:2001zs}.
\end{itemize}

For certain experiments the base CLAS system was complemented with ancillary 
detectors. For example, the measurement of the Deeply Virtual Compton 
Scattering (DVCS) process ($eH \rightarrow e' H' \gamma$, where $H$ is a 
nucleon or nucleus) necessitates an upgrade of the photon detection system.  
Indeed, with a 6 GeV electron beam, the majority of DVCS photons are produced 
at very forward angles, where the acceptance of the EC was poor. To extend the 
detection range, an inner calorimeter (IC) was built for the E01-113 experiment 
in 2005~\cite{Girod:2007aa}. The IC was constructed from 424 lead-tungstate 
(PbWO$_{4}$) crystals, covering polar angles between 5$^{\circ}$ and 
15$^{\circ}$~\cite{Hyon-suk}. To protect the CLAS detector and the IC from the 
large flux of the low energy M{\o}ller electrons, a 5~T solenoid magnet was 
placed around the target to shield the detectors.  To detect recoiling $\alpha$ 
particles from the coherent DVCS on Helium, a new radial time projection 
chamber (RTPC) was developed to track low energy nuclear fragments. The 
solenoid field was used to bend tracks and measure momentum of particles in the 
RTPC. The CLAS detector supplemented with both IC and RTPC was used in 2009 during a 
three months experimental run~\cite{proposal1,proposal2}
with a longitudinally polarized, 130~nA and 6.064 GeV electron beam
incident on a gaseous $^{4}$He target.

The original design of the RTPC was developed for the BoNuS 
experiment at Jefferson Lab which took data with CLAS in 
2005~\cite{Fenker:2008zz}. Significant improvements were made to the RTPC mechanical 
structure and fabrication technique that both increased the acceptance and 
reduced the amount of material in the path of the outgoing particles. 
Moreover, the data acquisition electronic was improved to increase the event 
readout rate. The enhanced design, 
used in the 2009 DVCS experiment, is described in section \ref{sec_design} of 
this article. The data acquisition system is described in section \ref{sec_readout}, 
the calibration methods in section \ref{sec_calib} and the tracking algorithm in section
\ref{sec_tracking}. Finally, the overall performances
of the RTPC are described in section \ref{sec_perfor}.

\section{RTPC design} \label{sec_design}

With a 6 GeV incident electron energy, the recoiling $^{4}$He nuclei from 
coherent DVCS have an average momentum around 300~MeV/c (12~MeV kinetic 
energy). Such low energy $\alpha$ particles are stopped very rapidly, so the 
RTPC was designed to be as close as possible to the target and fit inside the 
230~mm diameter shell and cryostat wall of the solenoid magnet bore of CLAS.

\begin{figure}[tb]
\centering
\includegraphics[{trim={0 12cm 0 0},clip, scale=0.4}]{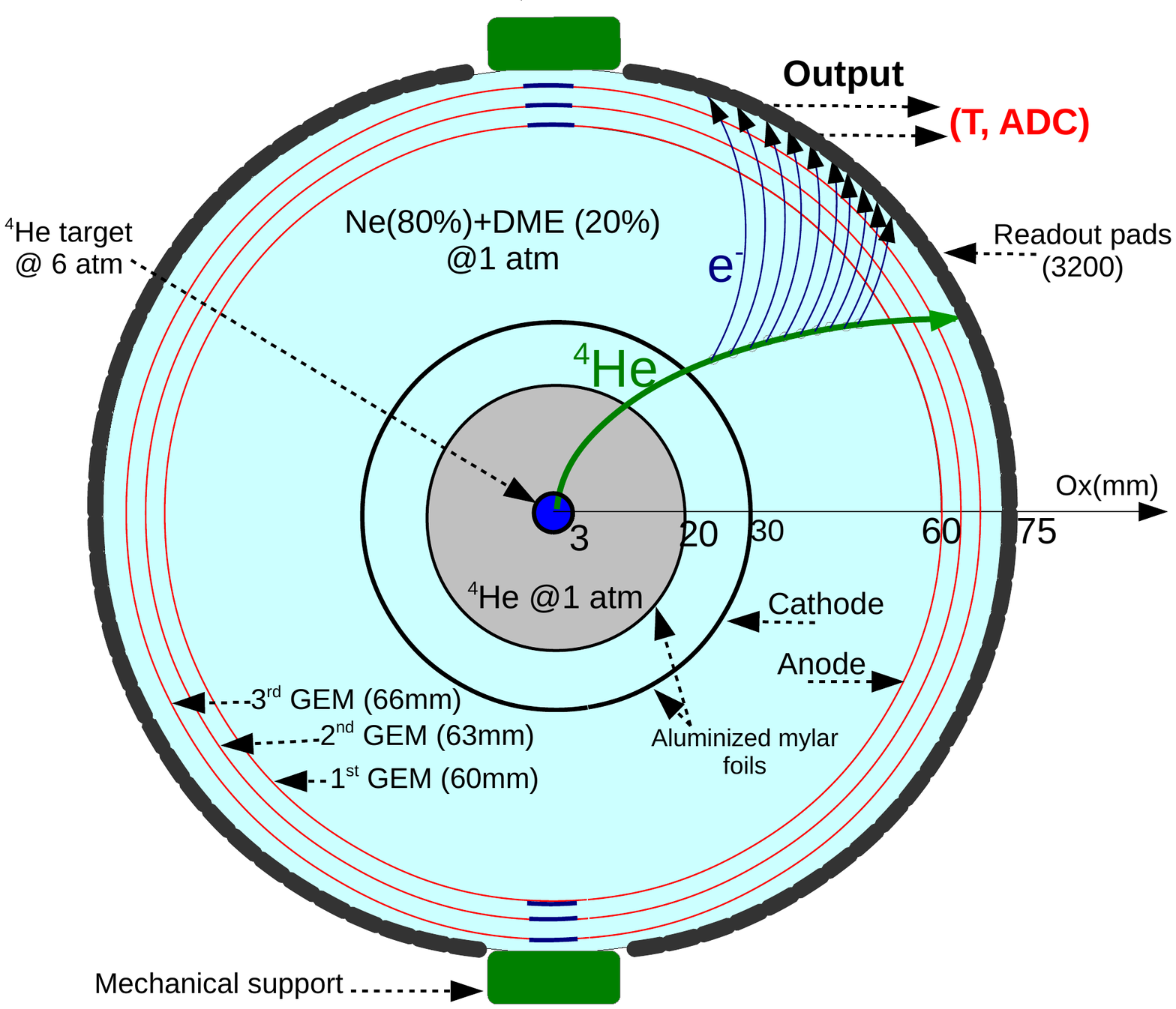}
\caption{Schematic drawing of the CLAS RTPC in a plane perpendicular to the beam 
direction. See text for description of the elements.} \label{fig:RTPC_1_4}
\end{figure} 

The new CLAS RTPC is a 250~mm long cylinder of 158~mm diameter, leaving just enough 
room to fit pre-amplifiers between the RTPC outer shell and the solenoid. The 
electric field is directed perpendicularly to the beam direction, 
such that drifting electrons are pushed away from the beam line. These electrons 
are amplified by three layers of semi-cylindrical gas electron multipliers 
(GEM)~\cite{Sauli:2016eeu} and detected by the readout system on the external 
shell of the detector as illustrated in Figure~\ref{fig:RTPC_1_4}. The RTPC is 
segmented into two halves with independent GEM amplification systems that cover 
about 80\% of the azimuthal angle.

We detail here the different regions shown in Figure~\ref{fig:RTPC_1_4} 
starting from the beam line towards larger radius:\\
\begin{itemize}
  \item The 6~atm $^4$He target extends along the beamline forming the detector 
     central axis. It is a 6~mm diameter Kapton straw with a 27~\textmu{}m wall 
     of 292~mm length such that its entrance and exit 15~\textmu{}m aluminum windows 
     are placed outside of the detector volume. The detector and the target are 
     placed in the center of the solenoid, 64~cm upstream of the CLAS center.
   \item The first gas gap covers the radial range from 3~mm to 20~mm. It is 
      filled with $^{4}$He gas at 1~atm to minimize secondary interactions from
      M\o{}ller electrons scattered by the beam. This region is surrounded by 
      a 4 \textmu{}m thick window made of grounded aluminized Mylar.
   \item The second gas gap region extends between 20~mm and 30~mm and is 
      filled with the gas mixture of 80$\%$ neon (Ne) and 20$\%$ dimethyl ether 
      (DME).  This region is surrounded by a 4~\textmu{}m thick window made of 
      aluminized Mylar set at $-4260$~V to serve as the cathode.
   \item The drift region is filled with the same Ne-DME gas mixture and 
      extends from the cathode to the first GEM, 60~mm away from the beam axis.  
      The electric field in this region is perpendicular to the beam and 
      averages around 550~V/cm.
   \item The electron amplification system is composed of three GEMs located at 
      radii of 60, 63 and 66~mm. The first GEM layer is 
      set to $\Delta V=1620$~V relative to the cathode foil and then each subsequent 
      layer is set to a lower voltage relative to the previous to obtain a 
      strong ($\sim$1600~V/cm) electric field between the GEM foils. A 275~V bias is 
      applied across each GEM for amplification.
   \item The readout board has an internal radius of 69~mm and collects charges 
      after they have been multiplied by the GEMs. Pre-amplifiers are plugged 
      directly on its outer side and transmit the signal to the data 
      acquisition electronics.
\end{itemize}

\begin{figure}[tbp]
\centering
\includegraphics[scale=0.70]{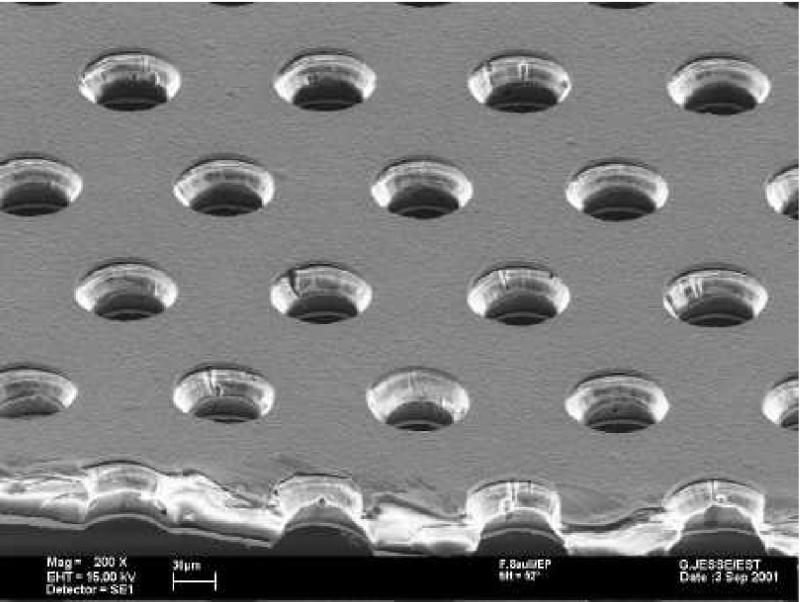}
\caption{Image of a typical GEM foil similar to the one used for our RTPC 
\cite{Sauli:2016eeu}.} 
   \label{fig:GEMs}
\end{figure}

The GEM technology has been chosen for the flexibility of the GEM foils,
which can be easily used to produce a curved amplification surface. Also, GEMs 
are known to have relatively low spark rate~\cite{Bachmann:2000kj}, which is 
important when trying to detect highly ionizing slow nuclei that deposit large 
amount of energy. The GEMs for this RTPC are made from a Kapton insulator 
layer, 50~\textmu{}m thick, sandwiched between two 5~\textmu{}m copper 
layers\footnote{The GEM foils were produced by Tech-Etch, Inc.}. The mesh of 
each GEM layer is chemically etched with 50~\textmu{}m diameter holes with 
double-conical shapes as illustrated in Figure~\ref{fig:GEMs}. The potential 
difference applied between the two copper layers of the GEM creates a very 
strong electric field in each hole leading to high ionization and 
amplification. 

The drift gas used in the experiment is a 80\%-20\% Ne-DME mixture. This choice 
has been made in order to balance the energy deposit, which is critical
for proper particle identification, with a reasonable
Lorentz angle. Calculations using the MAGBOLTZ program~\cite{Biagi:1999nwa} 
showed that with the 5~T solenoidal magnetic field, we would have a Lorentz 
angle of about 23$^\circ$ with this gas mixture.

The main structural improvement compared to the 2005 RTPC~\cite{Fenker:2008zz} 
was to obtain a better support structure for the GEM foils. In the previous version,
the RTPC was essentially composed of two independent half-cylinder separated
by their own structures. In this design, the installation of the GEMs was not 
very practical and wrinkles were visible on the GEM surfaces. To allow for a 
better installation and, at the same time, keep the mechanical structure out of 
the drift volume, the present design is based on self supporting GEM cylinder.  
We realized these cylinders by using fiber glass rings glued to each end of the 
GEM foils in order to install them independently in the RTPC after gluing and 
soldering operations. The rigidity of the GEM foils was enough for the 
structure to be self-supporting and only the upstream end of the cylinder was 
fixed to the mechanical support structure. This design only left a light 
fiberglass ring in the downstream end, reducing to a minimum secondary 
interactions.

\section{Readout System} \label{sec_readout}

\begin{figure}[tb]
   \centering
   \includegraphics[scale=0.55]{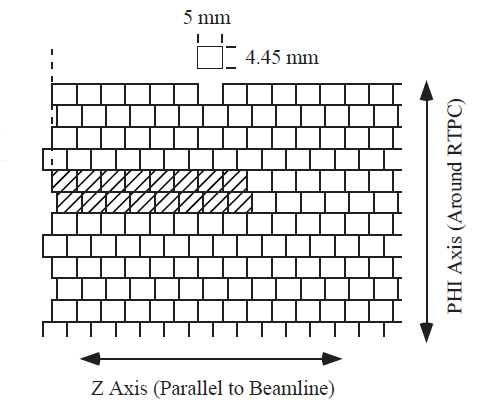}
   \caption[]{A schematic representation of the readout pads. The 
   shaded sixteen pads are a group of pads that are connected to the same 
pre-amplifier.} \label{fig:PADs}
\end{figure}

The RTPC electron collection system had 3200 readout pads. These elements were
located at the end of the amplification region, 69~mm from the central axis.
Figure \ref{fig:PADs} illustrates the configuration of the 5 by 4.45~mm pads,
where the shift between the rows was implemented to reduce aliasing. Each half 
of the RTPC had 40 rows and 40 columns of pads. The shaded region in 
Figure~\ref{fig:PADs} shows how pads were grouped to 16 channels pre-amplifier 
boards. The pre-amplifier boards, already employed in the BoNuS 
RTPC~\cite{Fenker:2008zz}, serve the dual purpose of inverting the RTPC signals 
polarity -- from negative to positive -- to match the requirements of the 
subsequent readout system, and driving signals through the 6~m long ribbon 
cable that connects to the readout system.

The readout system was an upgraded version of the original BoNuS RTPC 
system~\cite{Fenker:2008zz}, based on ALICE-TPC front end electronic boards 
(FECs)~\cite{ALICE-FEE}.  Each FEC hosted 128 channels, providing amplification 
and digitization of the input signals. For each event, 100 samples/channel were 
digitized by the ALTRO ASIC~\cite{EsteveBosch:2003bj}, operated at 10~MHz 
sampling frequency. In order to reduce the data size, the system was operated 
in zero-suppression mode, keeping data from $N_{PRE}=3$ samples before 
threshold crossing on the rising edge to $N_{POST}=3$ samples after threshold 
crossing on the falling edge.  The threshold level was set on each channel just 
above the noise level.

A new custom backplane was developed to connect FECs to the Readout Control 
Unit (RCU)~\cite{RCU}, allowing fast (200 MB/s) communication between the 
boards. The RCU board, used to distribute the trigger signal to different FECs 
and read data, was equipped with a fast optical link for data transferring to 
the main event builder. These features, together with the ``block-transfer'' 
readout mode making use of the FECs multi-event buffer, allowed to reach a 
significantly higher readout-rate compared to the original BoNuS system. During 
the 2009 run, the system was successfully operated with a DAQ rate of 3.1 kHz 
and a live time of 70~\%, for a luminosity of about 
$10^{34}$~cm$^{-2}\cdot$s$^{-1}$ and a beam energy of 6.064~GeV, to be compared 
with the 500 Hz obtained with the first BoNuS detector at similar run 
conditions.

During data reconstruction, the acquired samples were processed to obtain, 
for each readout pad, the accumulated charge (ADC) and the pulse time (T). Since 
pulse time was obtained as the time-stamp of the first sample above threshold, 
referred to the trigger time, the resolution was equivalent to the ALTRO 
sampling time of 100~ns.

\section{Calibration} \label{sec_calib}

The timing information collected from each signal is used 
to infer the origin of the ionization electrons, from which we
reconstruct the trajectory of the initial particle. The strength of these 
signals, recorded in ADC units, is then used to reconstruct 
the deposited energy per unit of length ($\small{\frac{dE}{dX}}$) which, 
together with the momentum calculated from the trajectory, enabled the particle 
identification. 

In this section we will detail the methods used to calibrate the drift time,
drift paths and gains of the detector. The drift paths were initially
calculated using the MAGBOLTZ~\cite{Biagi:1999nwa} program, then refined using
data to account for variations of the run conditions. The initial MAGBOLTZ
calibration was improved through several iterations of the
data driven process described below, with each time an increasing number of tracks 
reconstructed in the RTPC justifying a new iteration. The figures presented in this section
are the ones obtained while performing the last iteration of this
calibration process. We assume 
cylindrical symmetry in the chamber for the calibration of the drift parameters, 
such that do not depend on the azimuthal angle $\phi$.

\subsection{Maximum Drift Time Parametrization}

\begin{figure}[tb]
\centering
\includegraphics[{trim={0 12cm 0 0},clip, scale=0.4}]{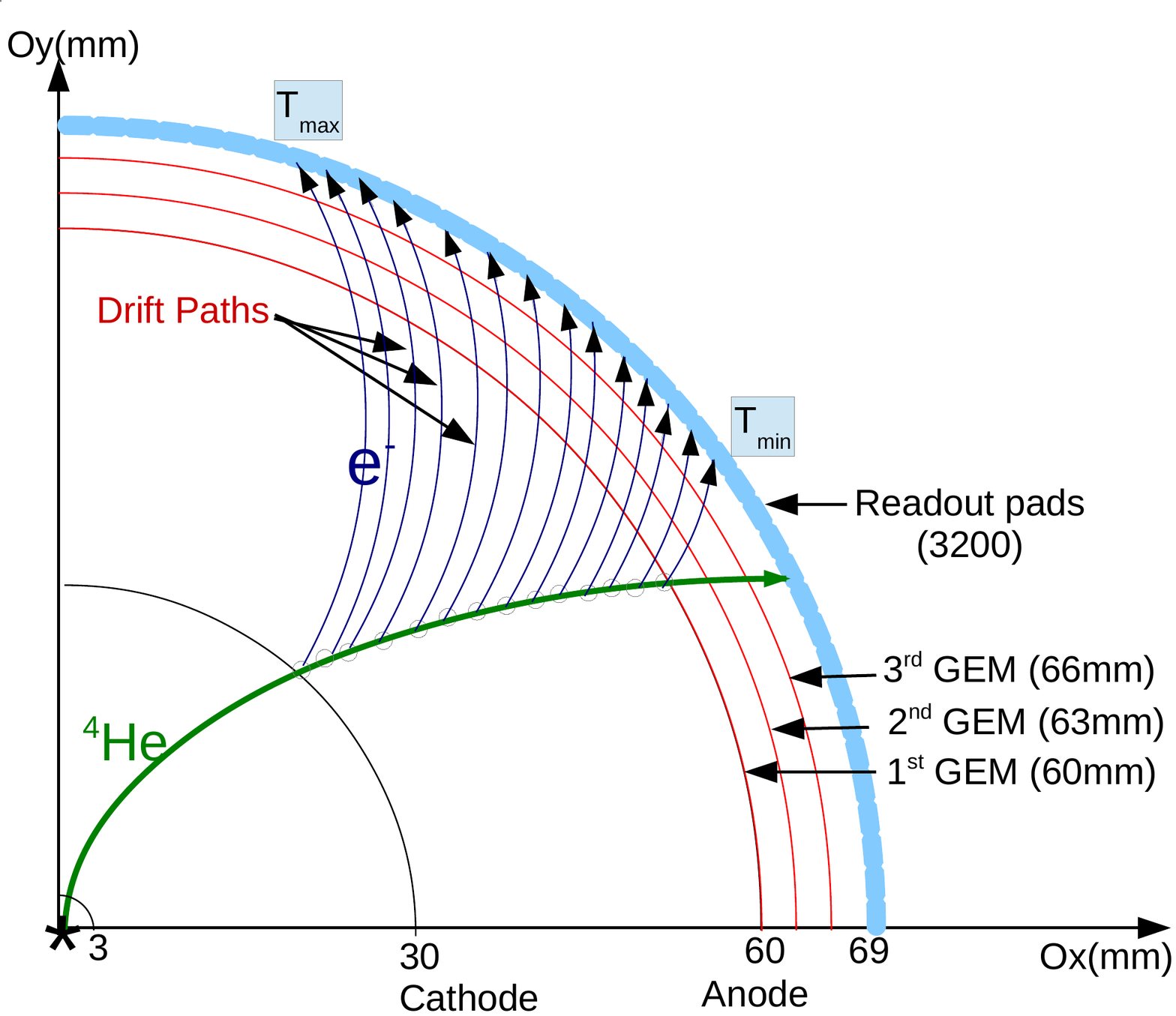}
\caption[]{A schematic drawing of a $^{4}$He track (in green) traversing the 
drift region, with the drift paths followed by the electrons (in blue). } 
\label{fig:RTPC_signals}
\end{figure}

The first step in the calibration is to fix the maximum drift time ($T_{max}$) that an
electron can arrive on the detection plane. This value is highly dependent on the gas 
mixture composition, as well as the electric and magnetic field experienced by the drift 
electrons. The maximum time corresponds to electrons released close to 
the cathode, which are the furthest from the readout pads as illustrated in 
Figure~\ref{fig:RTPC_signals}. Note that the minimum time is not calibrated, as we
use the trigger time for it ($T_{min}=15$), and that the time unit is the ALTRO 
sampling time of 100~ns.

\begin{figure}[t]
\centering
\includegraphics[scale=0.42]{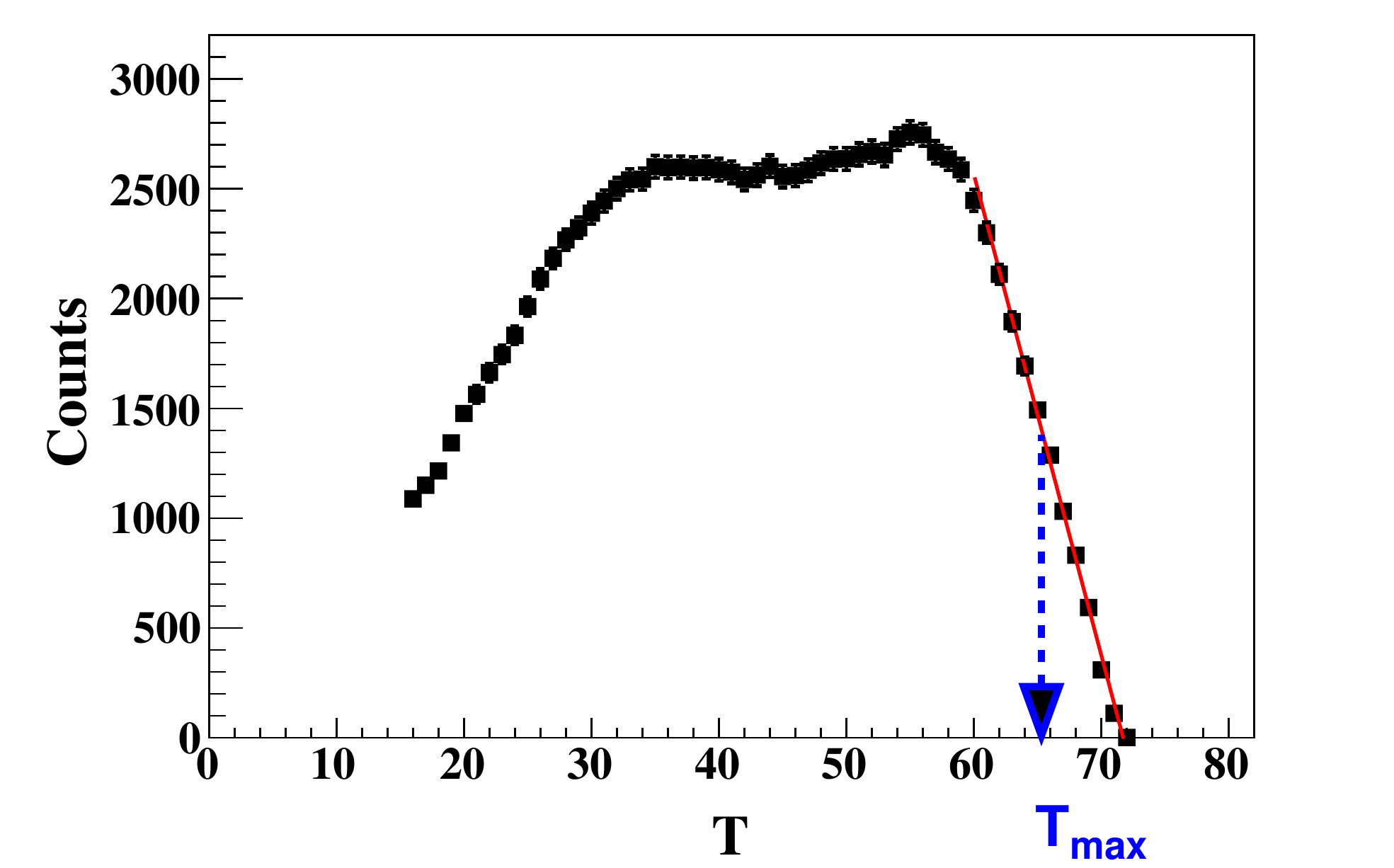}
\caption{Time distribution of the hits in one z-bin along the RTPC in one 
experimental run. } \label{fig:TDC_profile}
\end{figure}

\begin{figure}[t]
\centering
\includegraphics[scale=0.42]{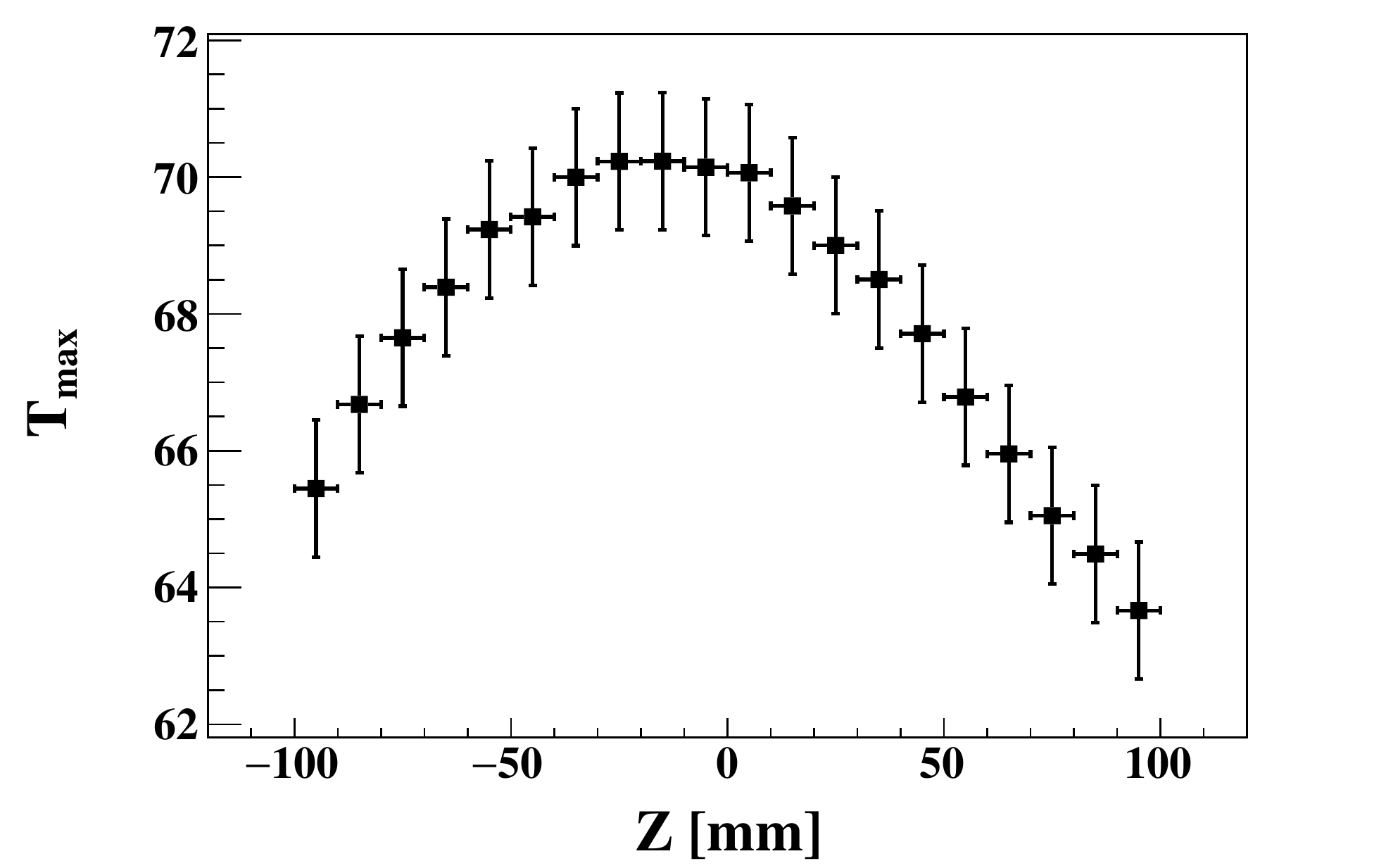}
\caption{Maximum time of collected hits as a function of the track
              position on the $z$-axis for one experimental run. } 
\label{fig:RunNumber_61551_TDCmax_Zslice}
\end{figure}

In order to experimentally determine the value of $T_{max}$, we analyze the
time profile of hits from identified good tracks, as presented in 
Figure~\ref{fig:TDC_profile}. We can clearly observe the expected dropping 
edge and define $T_{max}$, as the time at which the distribution passes 
below half the maximum number of hits in the histogram. This value was 
measured in bins along the 200~mm RTPC's length to take into account 
variations in the electric and magnetic field. We show this result for one 
experimental run in Figure~\ref{fig:RunNumber_61551_TDCmax_Zslice}. 

\begin{figure*}[t!]
\centering
\includegraphics[width=17.5cm]{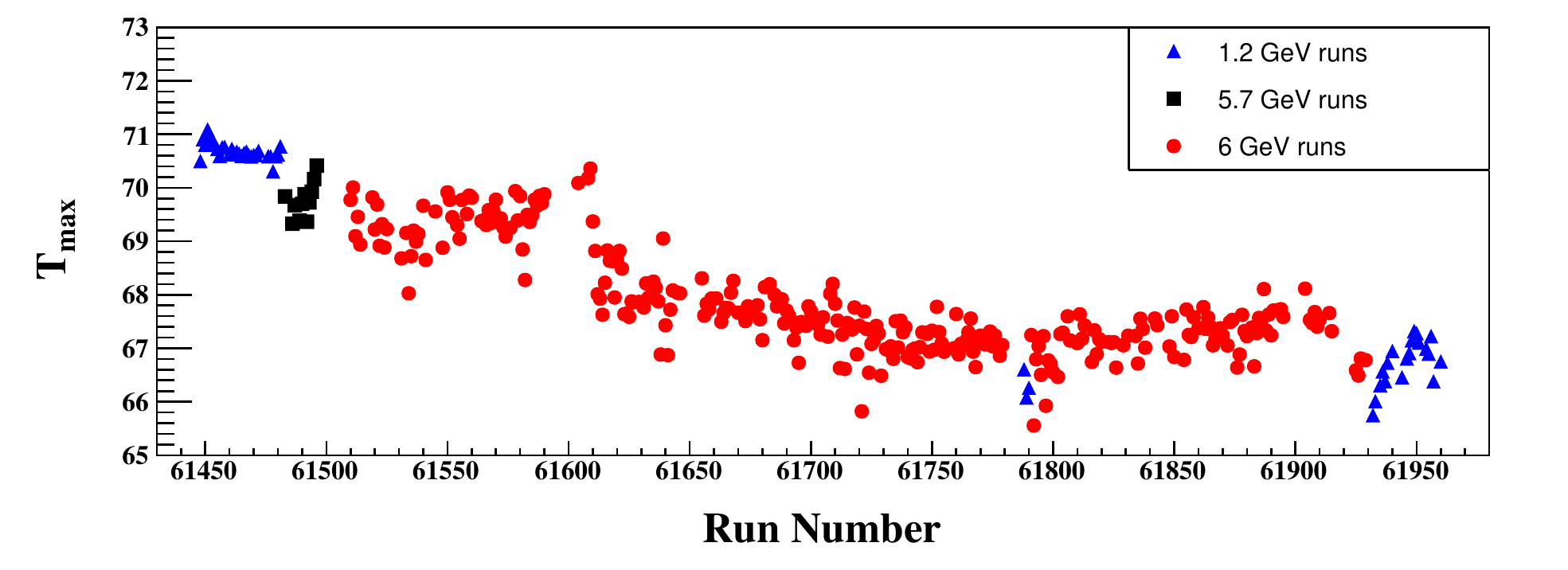}
\caption{$T_{max}$ versus the experimental run numbers.} 
\label{fig:Drift_run_number_1}
\end{figure*}

Figure~\ref{fig:Drift_run_number_1} shows the $T_{max}$ values for individual 
runs (approximately 2 hours long). We observe significant change of $T_{max}$
before and after run 61600, while variations within these periods are 
around 2\%. This is likely due to non-perfect experimental conditions, in 
particular possible contamination of our gas mixture. We indeed increase the gas flow 
around that time due to a small leak in the RTPC. While our 
gas system was kept slightly over atmospheric pressure to limit 
contamination from air or other external gases, it is likely that this leak was 
the source of modification of the drift velocity. 

To take this effects into account, we parametrized the maximum drift time 
as a function of both the position along the beam axis and the run number. These 
functions were extracted for the entire data set and implemented in the track 
reconstruction code for the next steps of calibration and track reconstruction.
   
\subsection{Drift Path Calibration}

The drift path is the trajectory followed by the electrons released through 
ionization in the gas. We initially calculated them with
MAGBOLTZ~\cite{Biagi:1999nwa}, but this calculation requires knowledge of the detector's 
geometry, of the gas mixture composition, and of the electric and magnetic 
fields over the whole volume of the detector. As we seen in the previous section, 
the conditions in the 
chamber were changing over time. Moreover, the 4 \textmu{}m foil used as a cathode 
is easily deformed, such that we expect the geometrical accuracy to be only of a few 
millimeters, also impacting our knowledge of the electric field. These 
problems, already encountered for the BoNuS RTPC 
calibration~\cite{Fenker:2008zz}, motivated the acquisition of specific 
calibration runs. These were taken with a lower energy electron beam (1.204 and 
1.269 GeV) to enhance the cross section of the elastic scattering 
($e^{4}$He$\rightarrow e^{4}$He). In this process, the measurement of
the electron kinematics allows to calculate the kinematics of the Helium nucleus. 
By comparing the calculated momentum and angle of the recoil alpha particle to the 
measurement in the RTPC, we tuned the drift paths independently of our 
knowledge of the chamber's conditions.

Based on the kinematics of the electrons in the calibration data, we generated 
the Helium nucleus in our RTPC GEANT4 simulation~\cite{GEANT4}. Then,
we compared the calculated GEANT4 trajectory of the Helium nuclei to 
the hits measured in the chamber. To perform the drift path extraction, 
we made a first approximation assuming a linear dependence between the radius 
of emission of the charge and its time of detection, and then refined our 
result accounting for the curvature of the drift path. The curvature was small 
enough, such that the process converged already on the second iteration.

\begin{figure}[t]
\centering
\includegraphics[scale=0.48]{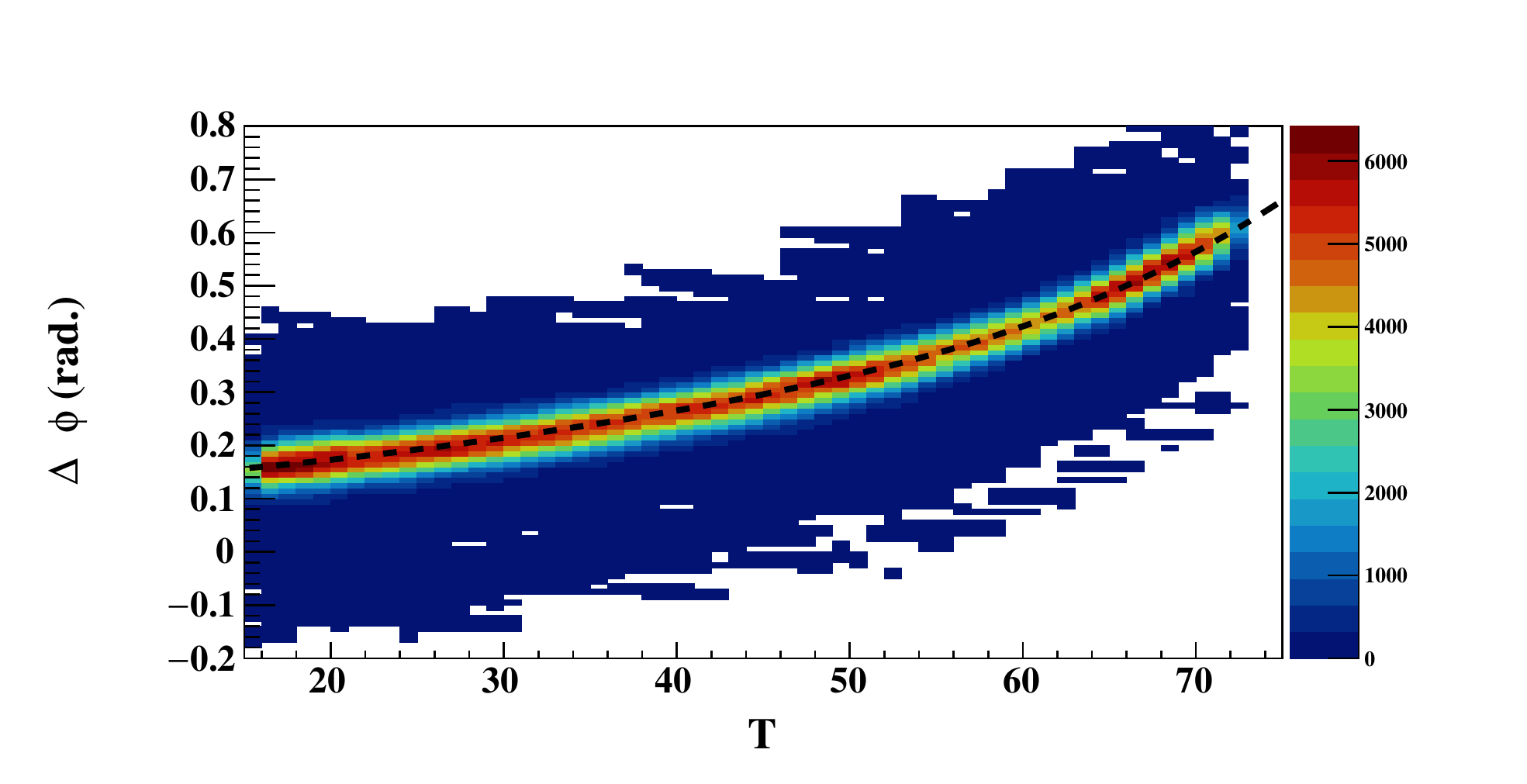}
\caption{$\Delta \phi$ versus T distribution for tracks
from one bin in longitudinal position along the RTPC. The black line represents 
our fit of the drift path in this bin.}
\label{fig:DELTA_PHI_TDC}
\end{figure}

At the end of the extraction procedure, the azimuthal difference between the 
detection pad and the ionization point ($\Delta\phi$) was extracted as a 
function of time. In Figure~\ref{fig:DELTA_PHI_TDC}, we show the resulting data 
points for one bin in z-coordinate of the RTPC, where the drift path is easily 
identified and eventually fitted for implementation in our reconstruction code.

To verify the stability of the drift paths, this procedure was carried out 
using both the 1.204 GeV data from the beginning of the run period and the 
1.269 GeV data from the end of the run period (shown in blue on 
Figure~\ref{fig:Drift_run_number_1}). We found very similar drift paths for the 
two data sets and concluded that any changes in the system only significantly 
affected the drift speed and thus the measured maximum time.

\subsection{Gain Calibration}

To calibrate the gains, we compared the experimental ADCs to the energy 
deposited for each pad individually in GEANT4 by similar simulated tracks 
(using the same elastic events as for the drift paths calibration). This 
requires a detailed simulation, including the drift paths and the diffusion of 
the charges along the path before reaching the pad to 
match the experimental data. We implemented in the GEANT4 simulation the
drift path we obtained from the calibration described above and implemented an {\it ad 
hoc} diffusion function to match the average number of hits recorded in the simulation with
the one from experimental data. To perform this step properly, we also implemented in
the simulation the details of the DAQ process to record hits. 

\begin{figure}[!t]
   \centering
\includegraphics[scale=0.42]{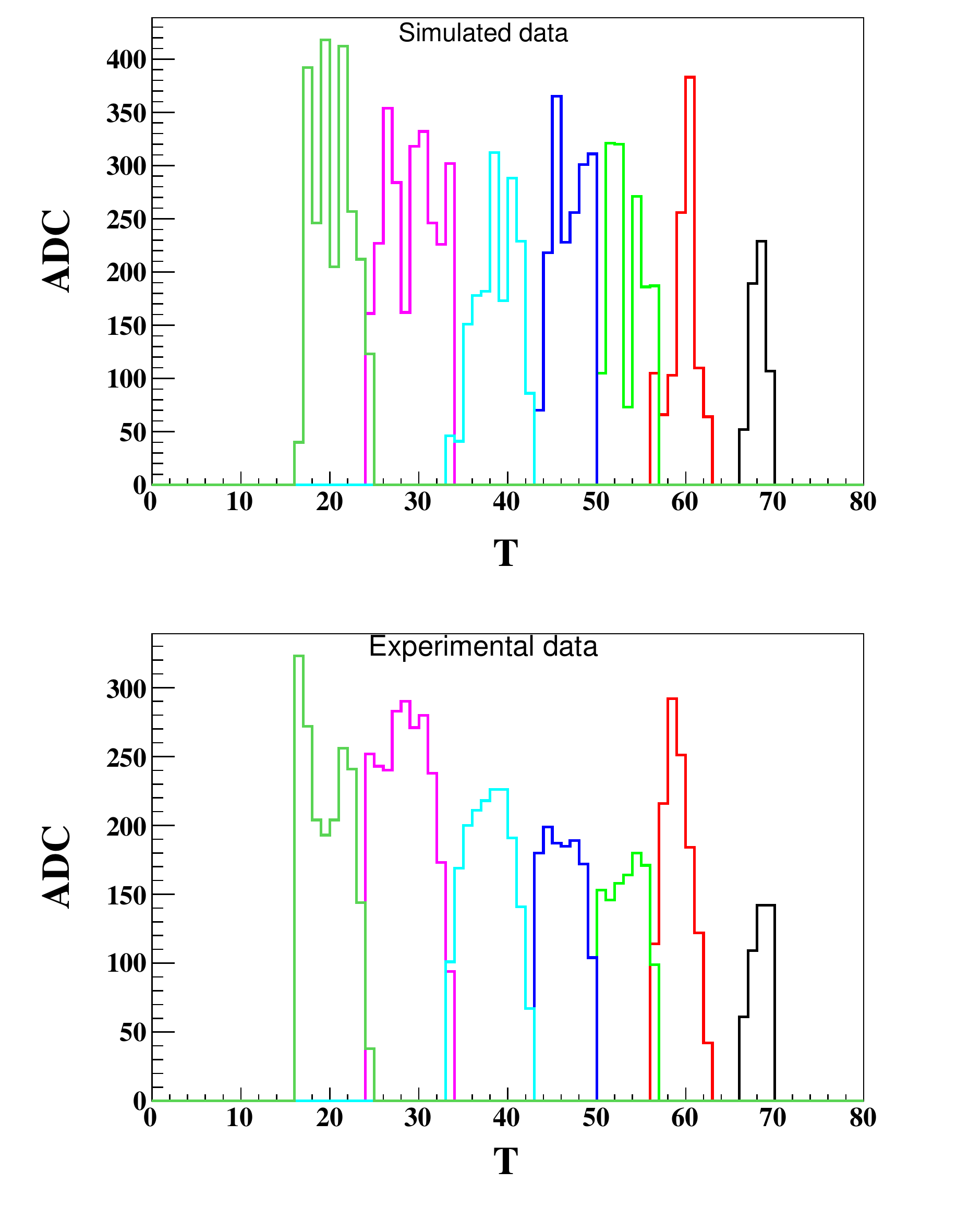}
\caption{Simulated (upper) and experimental (lower) ADC and T distributions 
of a single track. The different colors indicate the signals from different pads, 
the same color in the top and bottom figure indicate that the signal was registered
on the same pad.}
\label{fig:EVENT_adc_tdc}
\end{figure}

We then compared this simulation to experiment on an event by event basis as 
shown in Figure~\ref{fig:EVENT_adc_tdc}. The gain for each pad was calculated 
as the ratio of the measured ADCs to the simulated deposited energy. This 
provided our initial set of gains for each pads and it was applied to the 
experimental data. However, after this calibration some pads recorded lower ADCs 
than expected. So, we added a correction factor obtained by comparing  
the energy deposit on the pads within individual tracks. To do so, we compared, for 
each track, the energy deposit on a pad to the energy deposit in the whole track. 
The gain correction factors are extracted by averaging this ratio for a sample 
of elastic events and are thus completely independent from simulation. 

\begin{figure*}[!h]
\centering
\includegraphics[scale=0.73]{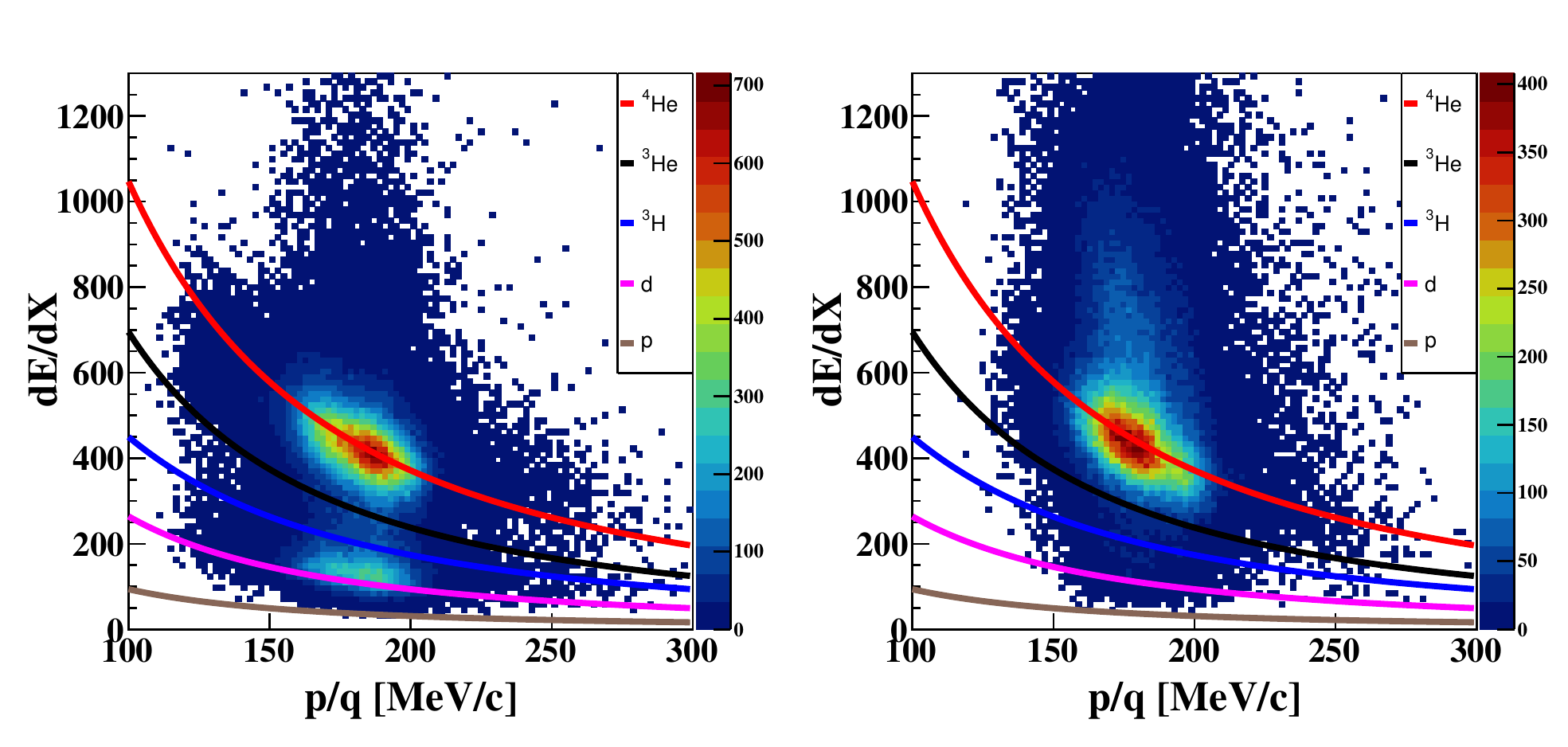}
\caption{$\small{\frac{dE}{dX}}$ vs. $p/q$ distributions for the left (on the 
   left) and for the right (on the right) half of the RTPC after gain 
   calibration. The lines are theoretical expectations from the Bethe-Bloch 
   formula for $^4$He 
   (red), $^3$He (black), $^3$H (blue), $^2$H (pink) and protons (gray).}
\label{fig:dedx_p_exp_2nd}
\end{figure*}

The results after calibration are shown in Figure~\ref{fig:dedx_p_exp_2nd}, where energy 
loss of particles is plotted against momentum over charge ratio. One can 
clearly see there the band for $^4$He in its expected position\footnote{The lower peak 
   in the left module of the RTPC comes from an unidentified problem in this half 
   of the detector that concerns 7\% of all the elastic events. 
   Our best guess is that these events are linked with a high voltage supply 
   issue, that would sometimes bring the GEM gain down. These events pass all 
   the elastic requirements and we found nothing that differentiates them from 
   other events but their low recorded ADCs. In particular, 94\% of pads
   are involved with these tracks such that we are sure they do not come from 
   a miscalibration of a part of the detector. Finally, we decided to discard
   these events for the calibration procedures.}.

\section{Track Reconstruction}
\label{sec_tracking}

\subsection{Noise Rejection}
Two independent noise signatures were identified in the raw data and removed 
in software prior to track reconstruction. Both are transient and isolated to 
a subset of the readout channels. 

\begin{figure}[tb]
   \centering
\includegraphics[scale=0.25]{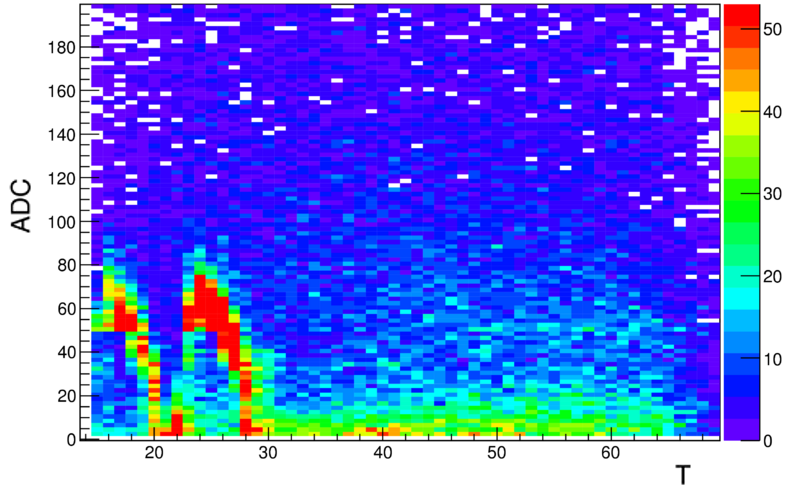}
\includegraphics[scale=0.25]{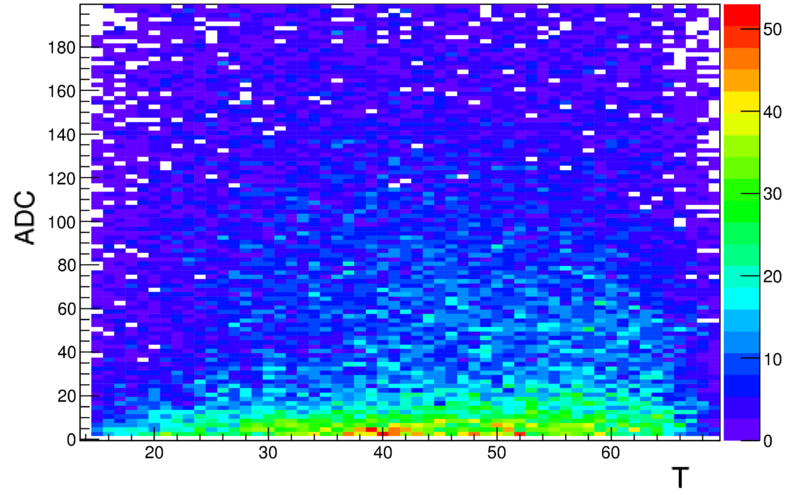}
\caption{The ADC vs. T spectrum for an example noisy channel before (top) and 
after (bottom) noise rejection algorithms.  Only hits associated with tracks 
are included, and the selection of events and tracks is the same in both 
plots.}
\label{fig:noise}
\end{figure}

The first is an oscillatory noise located early in the readout time window, 
shown in the top panel of Figure~\ref{fig:noise} for a particularly noisy 
channel. Its amplitude is similar to those of real tracks. About 18\% of the 
readout channels exhibit large contributions from this noise.  Due to its 
unique time-energy correlation, we use a pattern recognition algorithm and 
discard the hits coming from the channels that display this behavior on an 
event by event basis. The result of the procedure is illustrated in the bottom 
panel of Figure~\ref{fig:noise}.

The second noise signature was a coherent noise affecting about 25\% of the 
pre-amplifiers boards, when simultaneous hits in most of the 16 channels of the
board were recorded. An event-based technique to identify and remove this noise 
was developed based on counting simultaneous hits in each pre-amplifier group, 
and, if sufficiently large, perform a dynamic pedestal subtraction based on the 
average ADC of neighboring channels within this group.

The sources of these effects were not determined, but rejection techniques 
allowed to reconstruct 10\% more good tracks and recover 70 channels that were 
previously ignored due to excessive noise levels.

\subsection{Track Fitting}\label{sec_rec}

The tracking starts with reconstructing the spacial origin of the hits using 
the extracted drift speed and drift path parameters. For each registered hit, 
we calculate the position of emission from the signal time and the pad 
position. Then, chains of hits are created. The maximum distance between two 
close adjacent hits has to be less than 10.5 mm to chain them, which roughly 
corresponds to neighbors and next to neighbors. We fit the chains with a helix 
if they have a minimum of 10 hits. We then eliminate from the chain the hits 
that are 5 mm or farther from the fit, as they are not likely part of the same 
track. This new reduced chain is used for a second and final helix fit.

For energy deposition, the mean $\frac{dE}{dx}$ is calculated as
\begin{equation}
 \left\langle \frac{dE}{dX} \right\rangle= \frac{\sum\limits_{i} \frac{ADC_{i}}{G_i}}{L},
\end{equation}
where the sum runs over all the hits of the track, $G_{i}$ is the gain of 
the associated pad, and $L$ is the visible track length in the active drift 
volume. 

\subsection{Energy Loss Corrections}\label{sec_eloss}
Energy loss between the target and drift region was significant in the RTPC and 
necessitated a correction for optimal momentum reconstruction at the primary 
interaction vertex.  The dominant loss was in the 27~\textmu{}m thick Kapton target 
wall, with significant contributions also from the pressurized target gas and 
the foils before reaching the drift region.  Corrections were developed based 
on GEANT4 simulations with the full RTPC geometry and parametrized in terms of 
recoil curvature in the drift region and polar angle, separately for all recoil 
hypotheses ($p$, $d$, $^3$H, $^3$He, $^4$He).  At our average coherent $^4$He 
DVCS kinematics, energy losses were about 5 MeV, while for $e-^4$He elastic 
scattering losses were about 3 MeV, which corresponds to momentum corrections 
of 25\% and 15\%, respectively.

\section{Performance Studies}\label{sec_perfor}

\begin{figure}[tb]\centering
  \includegraphics[width=8cm]{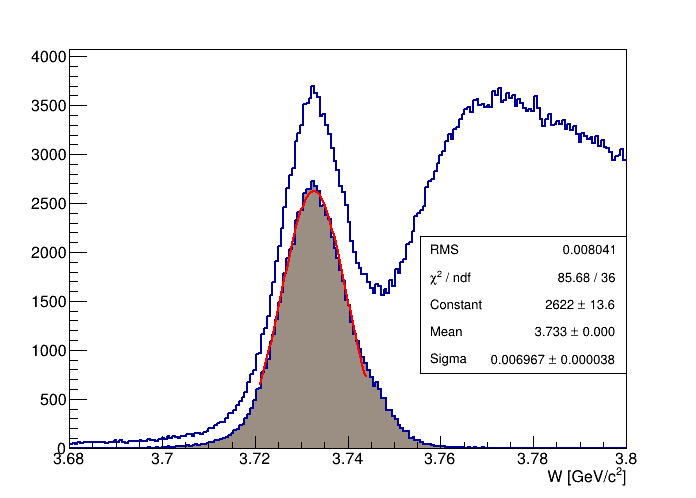}
  \caption{The recoil mass, $W$, distribution calculated from electron 
  kinematics before (``inclusive'') and after (``exclusive'') requiring a 
  matching track in the RTPC.\label{fig:w}}
\end{figure}

The primary data sample used for calibration and performance assessment of the
RTPC was elastic scattering with a 1.2 GeV electron beam. The electron momentum
and direction is measured with CLAS, which uniquely determines the expected recoiling
$^4$He kinematics. Matching requirements between reconstructed and expected $z$-vertex
and direction of the RTPC track provides a clean selection of elastically-scattered
$^4$He, shown in Figure~\ref{fig:w}.

\subsection{Resolution}

Elastic scattering was used to estimate the tracking resolution of the RTPC 
based on the residual between the expected and measured $^4$He tracks.  The RTPC
resolutions, after removing contributions from the electron, are shown in Table
~\ref{tab:reso}, and are very similar for the two halves of the RTPC.  Note that the
$\theta$- and $z$-resolutions are highly correlated.

\begin{table}[htbp]
\begin{center}
\begin{tabular}{|l|cccc|}
  \hline
& $\sigma_{z}$ &  $\sigma_{\theta}$ & $\sigma_{\phi}$ & $\sigma_{p}/p$\\
\hline
Left &  5.3 mm & 3.8$^{\circ}$ & 1.9$^{\circ}$ & 9$\%$ \\
Right & 6.5 mm & 4.0$^{\circ}$ & 1.9$^{\circ}$ & 8$\%$\\
\hline
\end{tabular}
\caption{The resolutions of the two modules of the RTPC for $z$-vertex, 
polar and azimuthal angles, and momentum.}
\label{tab:reso}
\end{center}
\end{table}

\subsection{Efficiency}

\begin{figure}[tb]
\centering
\includegraphics[width=8cm]{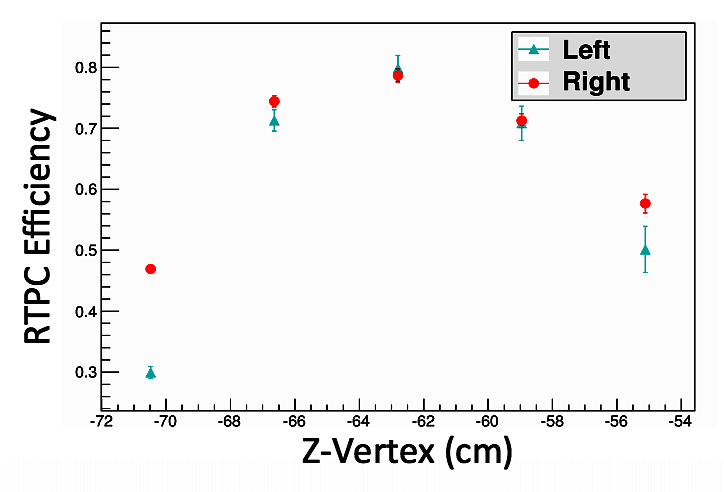}
\caption{The RTPC $^4$He detection efficiency as a function of the longitudinal 
   position along the detector.
 \label{fig:rtpc_eff}}
 \end{figure}

We measured the efficiency of the RTPC
using elastic scattering on $^4$He by comparing the inclusive yield, based
only on electron detection, to the exclusive elastic yield, where the Helium
recoil is also detected (see Figure~\ref{fig:w}). We present in 
Figure~\ref{fig:rtpc_eff} the results for the two halves of the detector. We 
observe that the left and the right modules have similar efficiencies except 
near the upstream target window. This difference is due to the large number of 
dead channels concentrated in this part of the left half of the detector

 \section{Conclusion}

We reported on the construction, operation and calibration of a small RTPC 
designed to measure $^4$He nuclei in high rate environment. The operation
of the detector was successful and allowed to detect Helium nuclei with a 75\%
efficiency and a readout rate
of 3.1 kHz triggered by the detection of high energy electrons and 
photons in the CLAS spectrometer. 

\section{Acknowledgments}

The authors thank the staff of the Accelerator and Physics Divisions at the 
Thomas Jefferson National Accelerator Facility who made this work possible.  
This work was supported in part by the French Centre National de la Recherche 
Scientifique (CNRS), the Italian Istituto Nazionale di Fisica Nucleare (INFN) 
and the U.S. Department of Energy. M.~Hattawy also acknowledges the support of 
the {\it Consulat G\'en\'eral de France \`a J\'erusalem}.  The Southeastern 
Universities Research Association operates the Thomas Jefferson National 
Accelerator Facility for the United States Department of Energy under contract 
DE-AC05-06OR23177. This material is based upon work supported by the U.S.  
Department of Energy, Office of Science, Office of Nuclear Physics, under 
contract number DE-AC02-06CH11357.

\end{document}